# Dense and Sharp Resonance Peaks in Stretched Moiré Waveguides

G. Alagappan* and C. E. Png

Institute of High-Performance Computing, Agency for Science, Technology, and Research (A-STAR), Fusionopolis, 1 Fusionopolis Way, #16-16 Connexis, Singapore 138632

*Corresponding author: gandhi@ihpc.a-star.edu.sg

**ABSTRACT**: In this article, we demonstrate dense resonant peaks in the transmission spectra of a rectangular waveguide inscribed with a stretched moiré pattern. We investigated an array of silicon waveguides with sinusoidally modulated cladding of varying depth of modulation. The investigation reveals a critical depth of modulation that splits the geometries into weakly scattering and strongly scattering regimes. Geometries in the weakly scattering regime resemble Bragg waveguides with shallow cladding modulation, whereas in the strongly scattering regime, the geometries resemble chains of isolated dielectric particles. The guided mode photonic bandgap for geometries in the strongly scattering regime is much larger than that of the weakly scattering regime. By inscribing stretched moiré patterns in the strongly scattering regime, we show that a large number of sharp peaks can be created in the transmission spectra of the waveguide. All periodic stretched moiré patterns can be identified with an $R$ parameter. The $R$ parameter indicates the ratio of the supercell period of the stretched system to the unstretched system. Our empirical study shows that the density of peaks linearly increases with $R$. The multiple resonance peaks evolve along well-defined trajectories with quality factor defined by exponential functions of $R$.

Moiré patterns [1] are periodic or quasi-periodic patterns that occur on a longer spatial scale. The longer scale super-periodicity destroys short-scale translational symmetries and induces coupling between various Bloch states of the original system, leading to rich physical phenomena and discoveries. In condensed matter physics, a moiré pattern is formed in a bilayer of graphene when one layer is twisted with respect to another layer [2-4]. Similar attempts have been made in the field of photonics using artificial bilayers of dielectric metasurfaces, where one metasurface is twisted with respect to the other. The collective works in this direction are currently evolving into a promising research field known as twisted photonics [5-9]. A new class of moiré system can be formed by stretching instead of the usual twisting. The topologies of these structures will consist of two periodic systems with the period of the second periodic system stretched with respect to the first one. Our prior works show that such structures fundamentally possess a unique spectral region in which a dense number of flat bands naturally appears [10-13]. Therefore, they have the great ability to revolutionize strong light-matter interactions by realizing a novel class of topologically-singular states [14-15] (including embedded eigenstates), frequency combs [16], broadband light localization [17], and efficient nonlinear devices [18].

Conventional periodic structures are single-period structures (SPS). They exhibit stopbands, which are spectral regions where wave propagation is forbidden. The physical principle behind this stopband formation is the Bragg resonance of the SPS, which causes waves with frequencies in the vicinity of the Bragg resonance frequency to experience a strong Bragg reflectivity and be unable to penetrate the bulk of the structure. The Fourier spectrum of the dielectric function of the SPS exhibits a strong fundamental harmonic at the spatial frequency, $G = 2\pi/a$, where $a$ is the period. Stretched moiré patterns can be formed by mixing two single periodic systems of slightly different periodicities, $a$ and $ra$, where $r$ is a number close to 1. In general, the mixing collapses all the translational symmetries existing in the periodic structure. However, long-range translational symmetries can be restored if $r$ is chosen to be a rational number. Consequently, in Fourier space, this will create two closely spaced harmonics: $G=2\pi/a$ and $G/r=2\pi/ra$. In real-space, a periodic structure of super-period $a_s = Ra$ is formed, where $R$ is the least integer multiple of $r$. The closest packing of $G$ and $G/r$ occurs when both of these spatial frequencies differ by one reciprocal lattice vector, $g=2\pi/a_s$ or equivalently $R=r/(r-1)$ [10-12].

When $r = 1$, we expect the two harmonics $G$ and $G/r$ to coincide and reproduce the physics of SPS. However, the condition $R = r/(r-1)$ imposes a topological singularity that prevents this resemblance. Instead of reproducing the physics of SPS, stretched moiré structures exhibit an exciting set of new physics in the limit $r \to 1$. Regardless of $r$, the stretched moiré topology exhibits spatial points where the two harmonics interfere destructively. In Ref. [10], it has been shown that one-dimensional (1D) Maxwell equations in the vicinity of destructive interference points mimic the standard equations of the quantum harmonic oscillator. The localized lights have been shown to display electric field envelopes defined by Hermite polynomials and Gaussian tails. In terms of the dispersion curve, 1D moiré structures exhibit a fully continuous curve in the limit $r \to 1$ [10,19]. This stands in direct contrast with the system of $r = 1$, where we have discontinuous dispersion (i.e., bandgaps). In high-index photonic crystal structures, the idea of stretching to create super-periodic moiré patterns and light localizations has been demonstrated by merging the dielectric functions of two photonic crystals with slightly different periodicities. Examples of such systems include 1D logically combined photonic crystals [11] and two-dimensional (2D) merged lattices [12].

Here, we present a study showcasing dense resonant peaks in the transmission spectra of a rectangular waveguide containing a stretched moiré pattern. Our investigation focuses on an array of silicon waveguides with cladding that has sinusoidal modulation of varying depth. The study uncovers a critical depth of modulation that divides the geometries into two regimes: weakly scattering and strongly scattering. In the weakly scattering regime, the geometries resemble a Bragg waveguide [20] with shallow cladding modulation, while in the strongly scattering regime, the geometries resemble a chain of isolated dielectric particles [21-22] . Geometries in the strongly scattering regime exhibit significantly larger guided mode photonic bandgaps compared to those in the weakly scattering regime. By introducing stretched moiré patterns in the geometries of the strongly scattering regime, and tailoring the parameter $R$, we demonstrate the creation of a large number of sharp peaks in the waveguide's transmission spectra. The multiple resonance peaks evolve along various well-defined trajectories with quality factors defined by exponential functions of $R$.

We started by considering the standard silicon waveguide embedded in silica with a height and width of $h = 220$ nm and $w = 500$ nm, respectively. This is a widely used single-mode waveguide in modern photonic integrated circuits [23-24]. Bragg cladding is introduced onto the waveguide by periodically patterning the width profile of the waveguide with a sinusoidal modulation, $w_{Bragg}(z) = w[1 + 4\varepsilon_1 \cos(Gz)]$ [**Figure 1(a)**], where $\varepsilon_1$ is the depth of the modulation. We consider a full range of depths of modulations, varying from $\varepsilon_1 = 0$ [straight waveguide without modulation] to the extreme case, $\varepsilon_1 = 0.25$ [i.e., the minimum of $w_{Bragg}(z)$]. The photonic band structures for various $\varepsilon_1$ are shown in **Figure 2**. These photonic band structures are calculated using the three-dimensional (3D) plane wave expansion method, and the bands are projected along the propagation direction, $z$. In these diagrams, the area above the light line is shaded, and the first two bands with *y-even* modes are shown in blue color. In the absence of Bragg modulation, the *y-even* modes represent the usual TE polarization modes of the silicon waveguide. The diagrams in **Fig. 2** show the evolution of the dispersion of this mode as a function of depth modulation. Near $\varepsilon_1 = 0$, the first two bands are simply the folded version of the dispersion curves of the waveguide. The two folded bands are degenerate when the wavevector, $k = \pi/a$. With the periodic width modulation, the continuous translational symmetry along $z$ is broken, and as a consequence, the folded bands couple to each other, and the degeneracy at $k = \pi/a$ is lifted. The evolution of frequencies as a function of $\varepsilon_1$ is shown in **Figure 3(a)** for the first two *y-even* bands. In this figure, the bandgap frequencies are shaded in red. When $\varepsilon_1 = 0$, we have degenerate bands with zero bandgap. Once $\varepsilon_1$ is increased from zero, we see the two bands separate, creating a bandgap. For $0.05 < \varepsilon_1 < 0.13$, the gap between the bands reduces, and a crossover of bands takes place at $\varepsilon_1 = 0.13$, yielding a zero bandgap. Thereafter, for $\varepsilon_1 > 0.13$, the trajectory of the bands deviates significantly, creating a huge bandgap, and thus, strong scattering. In **Fig. 3(b)**, we exhibit the transmission spectra of a finite Bragg waveguide with 41 periods for $\varepsilon_1 = 0$ to 0.25 in steps of 0.01. These transmission spectra are calculated using 3D finite difference time domain (FDTD) [25-26] method for $a$ = 300 nm. In the same figure, we show the band frequencies from **Fig. 3(a)**. As can be readily seen in **Fig. 3(b)**, both the plane wave expansion calculation of band frequencies and the FDTD calculation of transmission spectra were in good agreement. As expected, when $\varepsilon_1 = 0.13$, we observe scattering-less propagation. We define the region with $\varepsilon_1 < 0.13$ as a weak scattering region. This region occurs before the crossover and exhibits small bandgaps with shallow transmission dips. The band evolution in this region can be described as a mixing of two folded bands of a straight waveguide with a small perturbing Bragg cladding. The region $\varepsilon_1 > 0.13$ is defined as a strong scattering region. In the strong scattering region, we observe huge bandgaps with deep and wider transmission dips. The representative geometries from the weak scattering, scattering-free, and strong scattering regions are shown schematically in **Fig. 3(c)**. From this figure, we can see that in the weak scattering region, we have shallow Bragg claddings. In the strong scattering region, the modulations are stronger, and the resulting geometry mimics isolated dielectric nanoantennas. Therefore, the crossover $\varepsilon_1$ acts as a transition point from shallow Bragg cladding structures to strong antenna-like geometries.

Moiré patterns can be transferred to a silicon waveguide by modulating the width profile of the waveguide using two sinusoidal functions with spatial frequencies $G$ and $G/r$. The modulation function $w_{Bragg}(z)$ in Bragg waveguide needs to be replaced with $w_{moi}(z) = w\left[1 + 2\varepsilon_1 \cos(Gx) + 2\varepsilon_1 \cos\left(\frac{G}{r}x\right)\right]$. **Figs. 1(c) and 1(d)** show two examples of $w_{moire}(z)$ with $\varepsilon_1 = 0.1$ and $\varepsilon_1 = 0.23$. In **Figure 4**, we show scaled schematics of a series of stretched moiré waveguides with $R$ varying from 11 to 61. When $R$ increases, $r \to 1$, therefore with subsituition $r = 1$, we expect $w_{moire}(z) = w_{Bragg}(z)$. However, the condition $R = R/(r-1)$ imposes a singularilty that prevents the direct subsituition $r = 1$ for the limit $r \to 1$. From **Fig. 4**, we can see that regardless of $r$, the center of the stretched moiré waveguide is modulation-free. At the center, the two harmonics destructively interfere [see the expression of $w_{moire}(z)$]. This modulation-free region is protected from Bragg reflection and thus enables tight light localization [**10**]. Using 3D FDTD method, we calculated the transmission spectrum of moiré waveguide for full range of $\varepsilon_1$. **Figures 5(a), 5(b), and 5(c)** show arrays of transmission spectra for $R$ = 21, 31, and 41, respectively. For comparison, we also plotted the transmission spectrum of Bragg waveguides in the same figure (lines in orange color). From **Fig. 5**, we observe that shallow peaks and valleys (ripples) appear in the transmission spectra of Moiré waveguides when $\varepsilon_1$ falls within the weak scattering geometrical regime. On the other hand, in the strong scattering regime, we have pronounced deep peaks in the transmission spectra of Moiré waveguides. As $R$ increases, we can clearly notice that the sharpness of the peaks and their counts increase significantly.

In order to analyze the density of sharp peaks as a function of $R$ using 3D FDTD calculations, a significant computational effort is required. As $R$ increases, the size of the geometry, the number of resonant peaks, and their sharpness increase tremendously. This demands a huge computational domain and long calculation time. To circumvent this problem, we attempted an effective index approach [**27-28**] in which 3D calculations are reduced to 2D using an effective refractive index [$n_{eff}$]. In this method, the refractive index of the waveguide core in 2D is taken as the effective index of the 3D straight waveguide. This method works best for a narrow range of frequencies and shallow modulations in the weak scattering regime. For broadband computations in the strong scattering regime, the waveguide dispersion plays a central role. Broadband computations can be handled by calculating the effective indices as a function of wavelength, $n_{eff}(\lambda)$, and substituting the core material (with a refractive index of $n_c$) in 3D calculations with a dispersive material of refractive index $n_{eff}(\lambda)$ in 2D calculations. On the other hand, to address the strong perturbation to the waveguide geometry in the strong scattering regime, we conducted an optimization of $n_c$, where $n_c$ is varied to achieve a good agreement between 2D and 3D calculations. **Figure 6** displays the results of the calculation ($R = 21$, $\varepsilon_1 = 0.23$) for various $n_c$ values. As seen from this figure, a good match between 2D (blue lines) and 3D (red markers) is obtained for $n_c$ values between 3.65 and 3.7. Further optimization with fine values of $n_c$ yields an optimized $n_c$ of 3.68. Using this core index, we simulated an array of moiré waveguides with $\varepsilon_1 = 0.23$.

In **Figure 7(a)**, we plot the calculated transmission spectra for $R$ varying from 21 to 81, and in **Fig. 7(b)**, we show the locations of sharp peaks for various $R$ in the considered wavelength window. From **Fig. 7(b)**, it can be observed that the density of peaks increases almost linearly with $R$. For every increment of $R$ in steps of 10, an additional sharp peak appears in the wavelength window. **Fig. 7(b)** also clearly exhibits wavelength trajectories of multiple resonances as a function of $R$. The quality factors for the selected trajectories are given in **Fig. 7(c)**. In this figure, the circular markers represent numerically calculated quality factors, while the solid lines represent straight-line approximations. It is important to note that the vertical axis in **Fig. 7(c)** is logarithmic, showing that the quality factors explode exponentially as a function of $R$.

In the limit as $r \to 1$ (i.e., increasing $R$), instead of reproducing the band gaps of a single-period structure, we have a set of dense sharp resonances. This can be understood from our earlier work on ideal one-dimensional periodic moiré structures [**10-12**]. In moiré structures with finite $r$, only the longer-scale periodicity in the spatial dimension exists. This breaks the translational symmetry on the shorter spatial scale and consequently leads to a strong coupling of Bloch modes that are initially degenerate in the single-period system. The strong coupling of these modes induces a dense number of flat bands in a spectral region called the strongly coupled regime. As we have demonstrated here, in the transmission spectrum, the flat bands emerge as sharp transmission peaks.

In summary, we have demonstrated controllable dense sharp resonance peaks in the transmission spectra of a rectangular silicon waveguide inscribed with a stretched moiré pattern. We began with the analysis of the Bragg waveguide with various depths of modulation and showed that there exists a critical depth beyond which the bandgap enlarges significantly and scatters like a dielectric nanoantenna. By creating stretched moiré patterns with depths of modulation falling in this regime, we show that strong resonance can be generated at multiple wavelengths of light. The density of resonance peaks and their quality factors can be controlled by the $R$ parameter of the stretched system. As $R$ increases, the multiple resonance peaks evolve along various trajectories with exponentially increasing quality factors. In this article, to tackle the broadband computational complexity of large dielectric structures with high quality factors, we adapted an effective index approach in which three-dimensional computations are reduced to a two-dimensional space by simultaneously adjusting the dispersive effective refractive indices with the refractive index of the waveguide core material.

FIG. 1 G. ALAGAPPAN ET AL

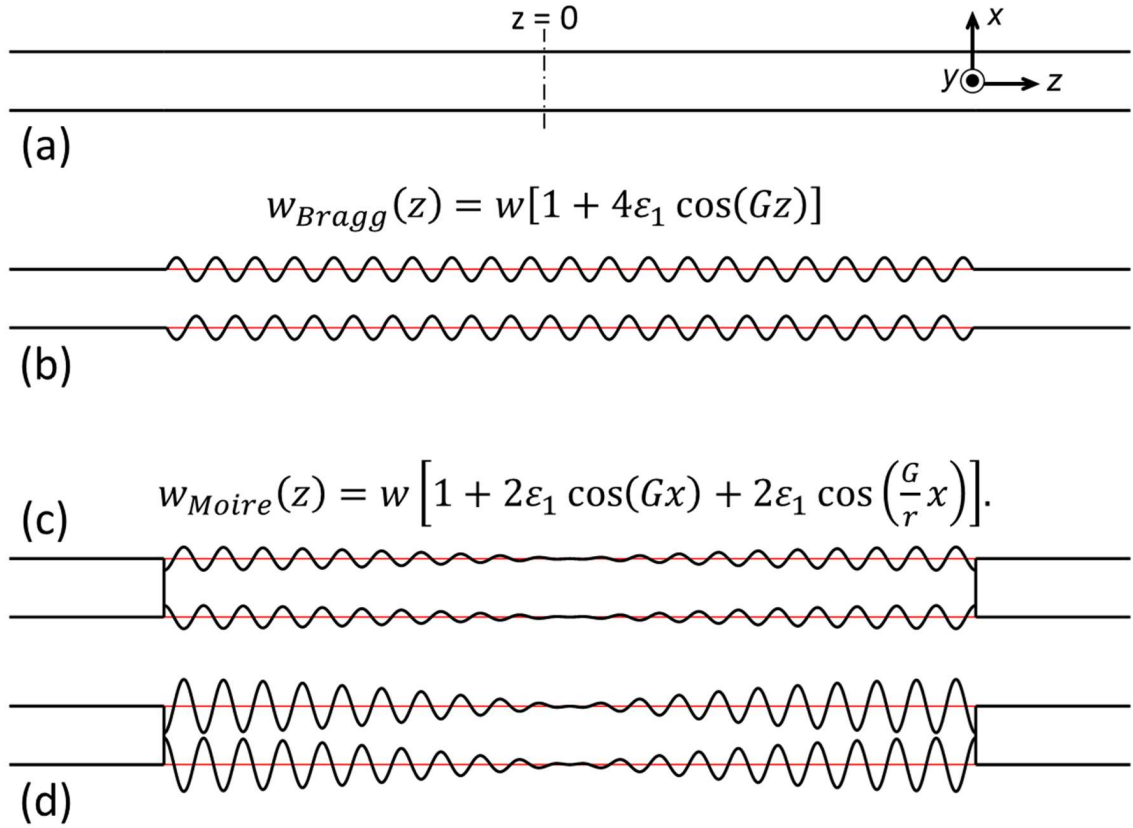

Fig. 1 Schematics of the various photonic waveguide geometries considered in this article. Planar views of (a) straight waveguide, (b) waveguide with periodically modulated width (Bragg waveguide), and waveguides with width profile define by moire pattern for (c) $\varepsilon_1 = 0.1$ and (d) $\varepsilon_1 = 0.23$.

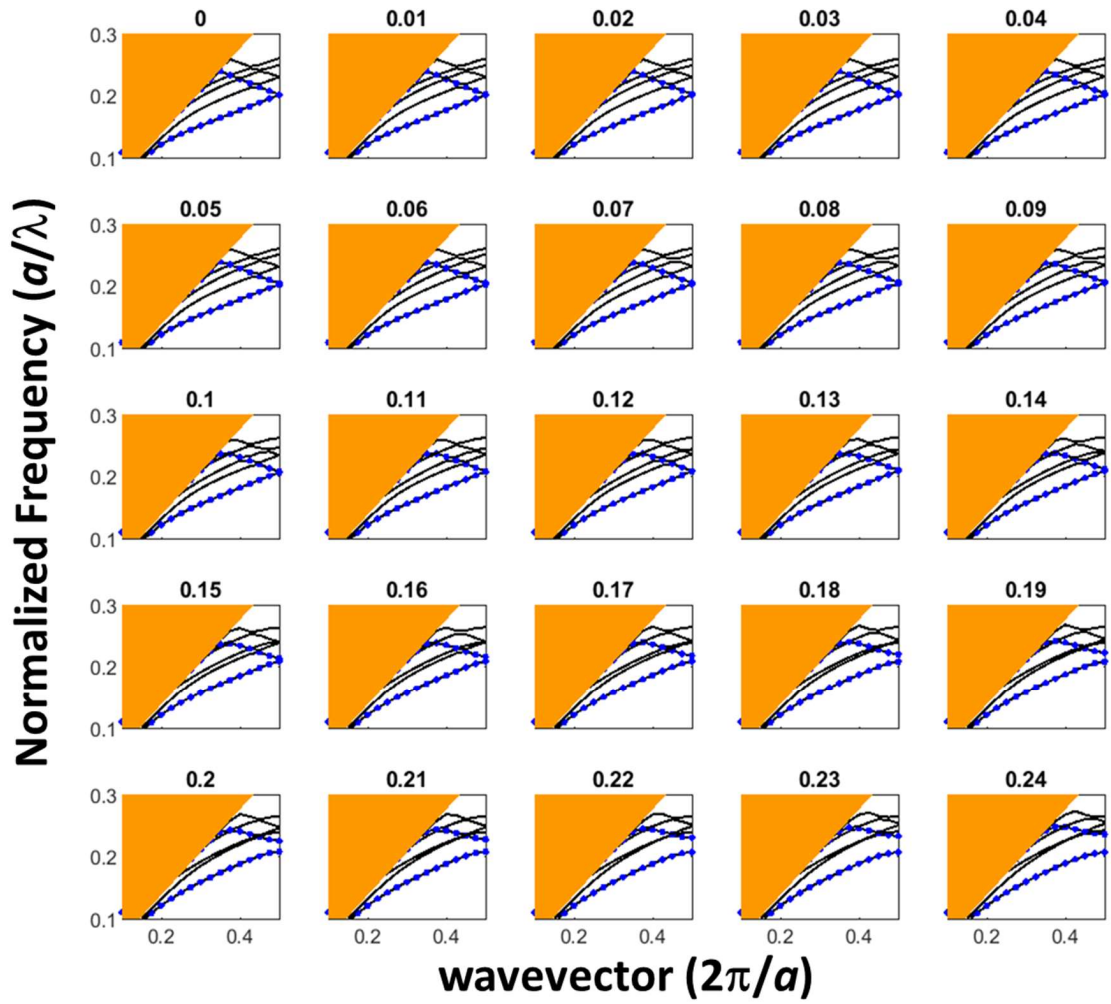

Fig. 2 Projected band structure of Bragg waveguide with various depth of modulations [$\varepsilon_1$] (indicated at the top of every plot). The first two *y*-even bands are shown in blue lines.

FIG. 3 G. ALAGAPPAN ET AL

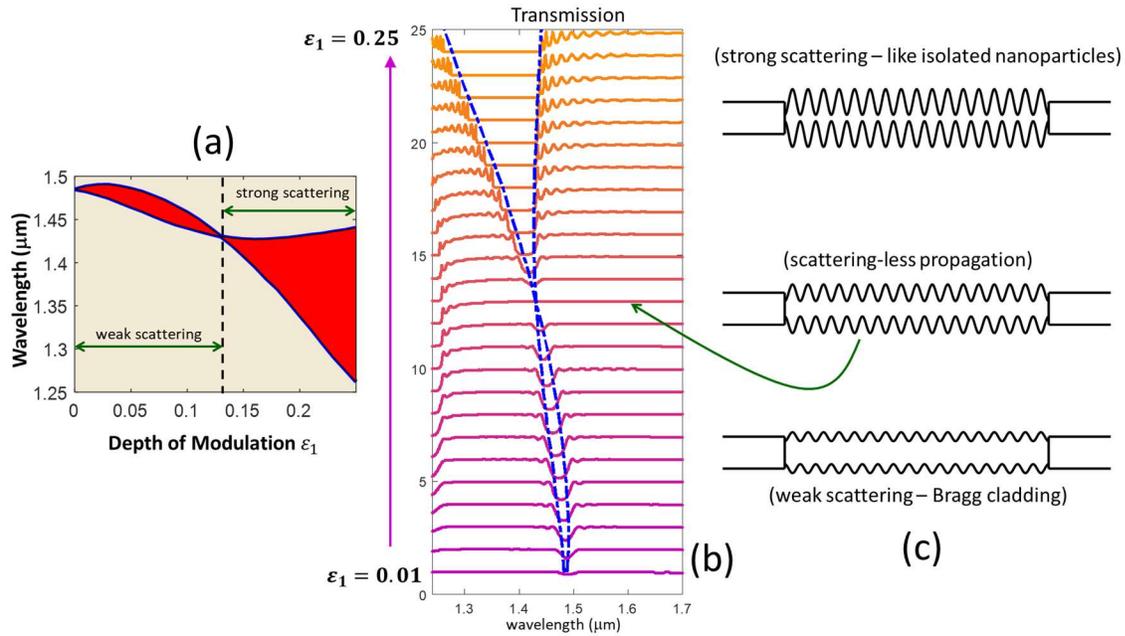

Fig. 3 Salient properties of Bragg waveguides. (a) Evolution of the first two *x*-even bands as a function of depth of modulation. The shaded red area represents guided mode photonic bandgap. (b) Transmission spectra of Bragg waveguide for $\varepsilon_1 = 0$ to 0.25 in steps of 0.01. Depth of modulations are increased from $\varepsilon_1 = 0$ to 0.25 in steps of 0.01. For every increment, the transmission is added one, cumulatively. (c) Representative waveguide geometries.

FIG. 4 G. ALAGAPPAN ET AL

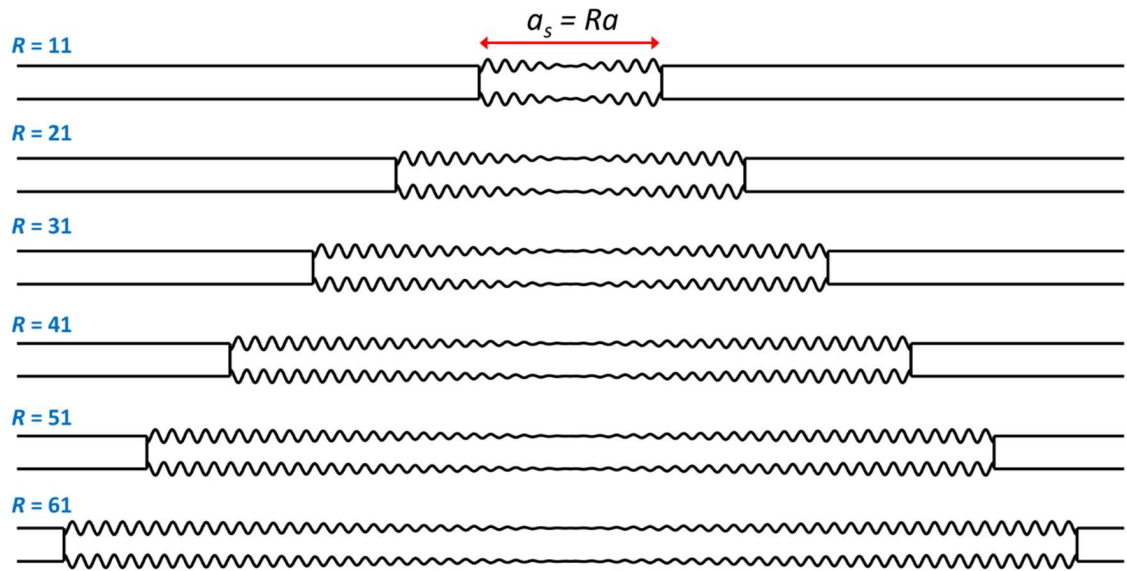

Fig. 4 Schematics of stretched moire waveguide with increasing $R$ or equivalently $r \to 1$. Regardless of $r$, the center of the stretched moiré waveguide is modulation-free.

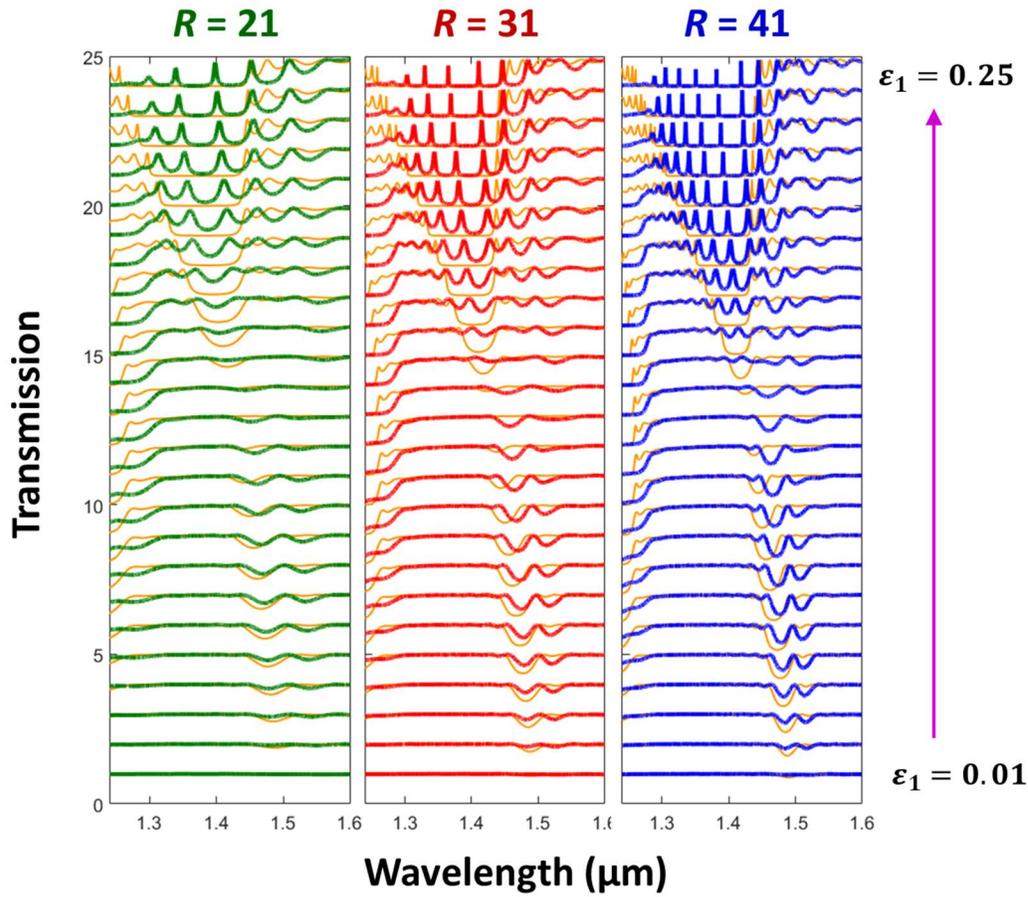

Fig. 5 Transmission spectra of stretched moire waveguide as a function of depth of modulations. In this figures, transmission spectra of the Bragg waveguides are shown in orange lines. The depth of modulations is increased from $\varepsilon_1 = 0$ to 0.25 in steps of 0.01. For every increment, the transmission is added one, cumulatively.

FIG. 6 G. ALAGAPPAN ET AL

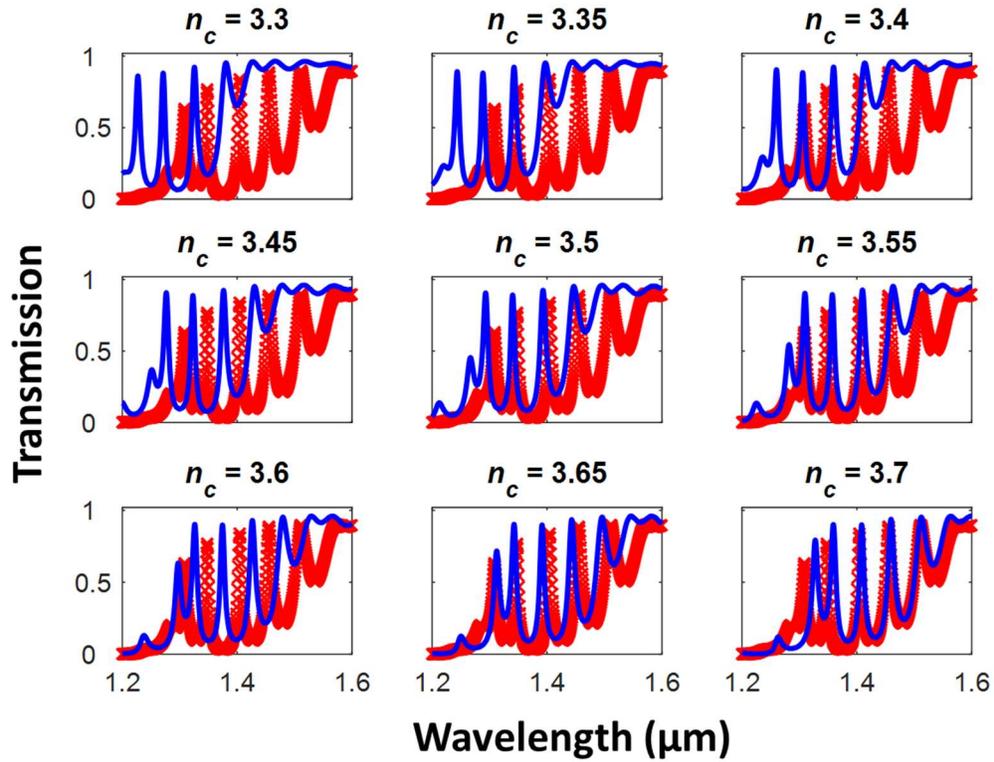

Fig. 6 Comparison between 2D FDTD with dispersive effective refractive index (blue line) and 3D FDTD (red line) methods. In these figures the refractive index of the waveguide core (indicated at the top of each plot) is varied to attain a good match between the two methods.

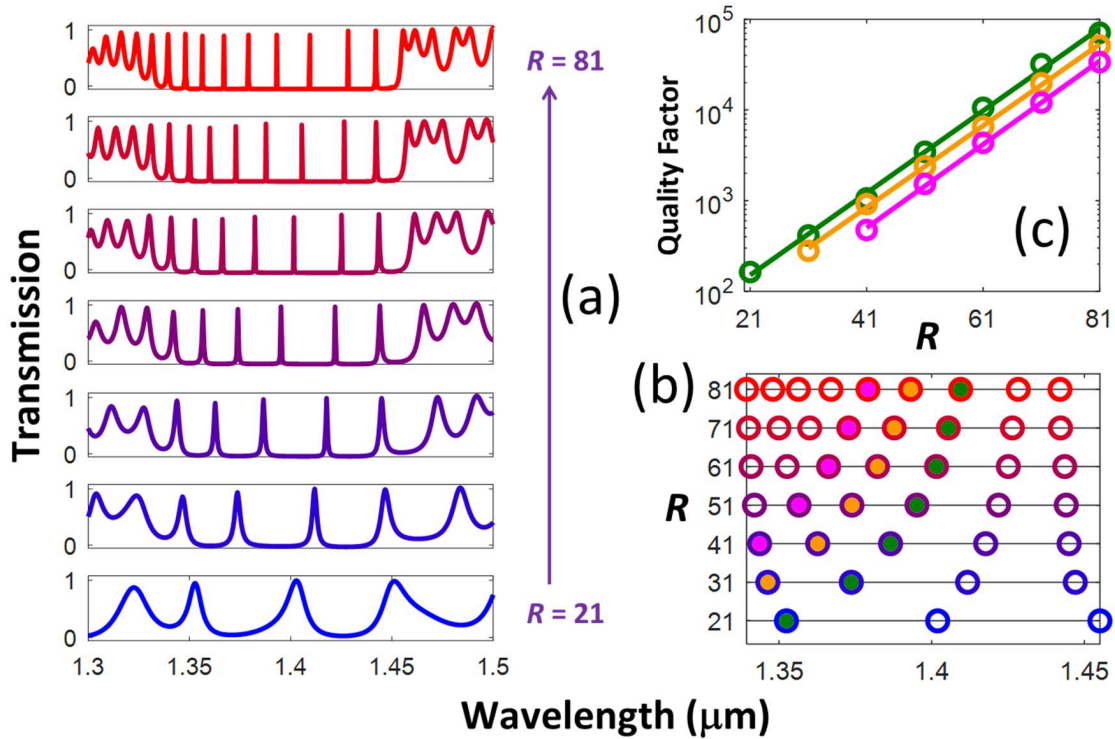

Fig. 7 Density and the sharpness of the peak as a function of the $R$ parameter. (a) Transmission spectra of stretched moire waveguide for $R$ = 21 to 81 (in steps of 10). (b) Resonance wavelengths [i.e., location of transmission peaks in (a)] for various $R$. Closed circles: Trajectories of resonance wavelength evolution. (c) Quality factor as a function of $R$ for trajectories in (b). Same trajectories are identified with the same color in both (b) and (c). In (b) open circles are numerically evaluated, and solid lines are linear fits.